\documentclass[aps,prl,twocolumn,superscriptaddress,floats,a4paper,preprint,tightenlines,10pt]{revtex4-2} 

\usepackage{epsfig}
\usepackage{amsmath}
\usepackage{graphicx}			
\usepackage{dcolumn}			
\usepackage{xcolor}				
\usepackage{hyperref}			
\usepackage[capitalize]{cleveref}		
\usepackage{amssymb,xcolor,graphicx}  	

\hypersetup{				
	colorlinks   = true, 	
	urlcolor     = blue, 	
	linkcolor    = blue, 	
	citecolor    = blue     
}
\crefformat{equation}{Eq.~#2(#1)#3}				
\crefformat{figure}{Fig.~#2#1#3}				
\crefformat{section}{Sec.~#2#1#3}				

\newcommand{\hide}[1]{}

\newcommand{\nbb}[1]{{\color{black} #1}} 		

\begin{document}
\title{Lenses in Minkowski space optimize queues by hiding bottleneck units}

\author{Eitan Bachmat}
\affiliation{Department of Computer Science, Ben-Gurion University, Israel}
\author{Sveinung Erland}
\email[Corresponding author: ]{sver@hvl.no}
\affiliation{Department of Maritime Studies, Western Norway University of Applied Sciences, Norway}
\author{Vidar Frette}
\affiliation{Department of Safety, Chemistry and Biomedical laboratory sciences, Western Norway University of Applied Sciences, Norway}
\author{Jevgenijs Kaupu\ifmmode \check{z}\else \v{z}\fi{}s} 
\affiliation{Institute of Technical Physics, Riga Technical University, Latvia}
\affiliation{Institute of Mathematical Sciences and Information Technologies, University of Liepaja, Latvia}


\date{\today}
\begin{abstract}
We mathematically construct focal lenses for positive-mass particles in Minkowski space. Particles emanating from a point source, refocus with synchronized proper time regardless of initial direction and velocity. Minkowski space offers a useful representation of the passenger queue formed during airplane boarding. The blocking relations between passengers coincide with the past-future relation in spacetime, and the boarding time is determined by the longest blocking chain.  
All blocking chains can be kept equally long if slow passengers are placed according to a lens in Minkowski space.
The two media of the lens correspond to a grouping of fast and slow passengers, and the lens specifies the optimal placement of fast and slow passengers in the queue at the gate.	
Not only the boarding time, but also its variance is minimized.
\end{abstract}
\maketitle

{\em Introduction.---}
Spatial dimensions and time combined into {\em spacetime}, first formulated
by Hermann Minkowski, has been a valuable tool in many branches of physics.  
We will demonstrate that lenses in Minkowski space are useful tools for optimization.  

In optics, light rays that pass through a {\em converging lens} meet at a focal point. The construction of a lens follows from Fermat's principle, which states that a ray follows the path that minimizes the travel time, and from this a lens can be defined by a Cartesian oval (see \cref{fig:lenses}(a)). In this Letter we construct corresponding lenses for positive-mass particles in Minkowski space (\cref{fig:lenses}(b)), which leads to optimal solutions of hard queue problems.
Thus, we fulfill part of Minkowski's program, where he envisioned spacetime as a crucial concept for understanding the entire physical world \cite{Galison:1979}.
{\begin{figure}[h]
		\includegraphics[height=0.5\linewidth]{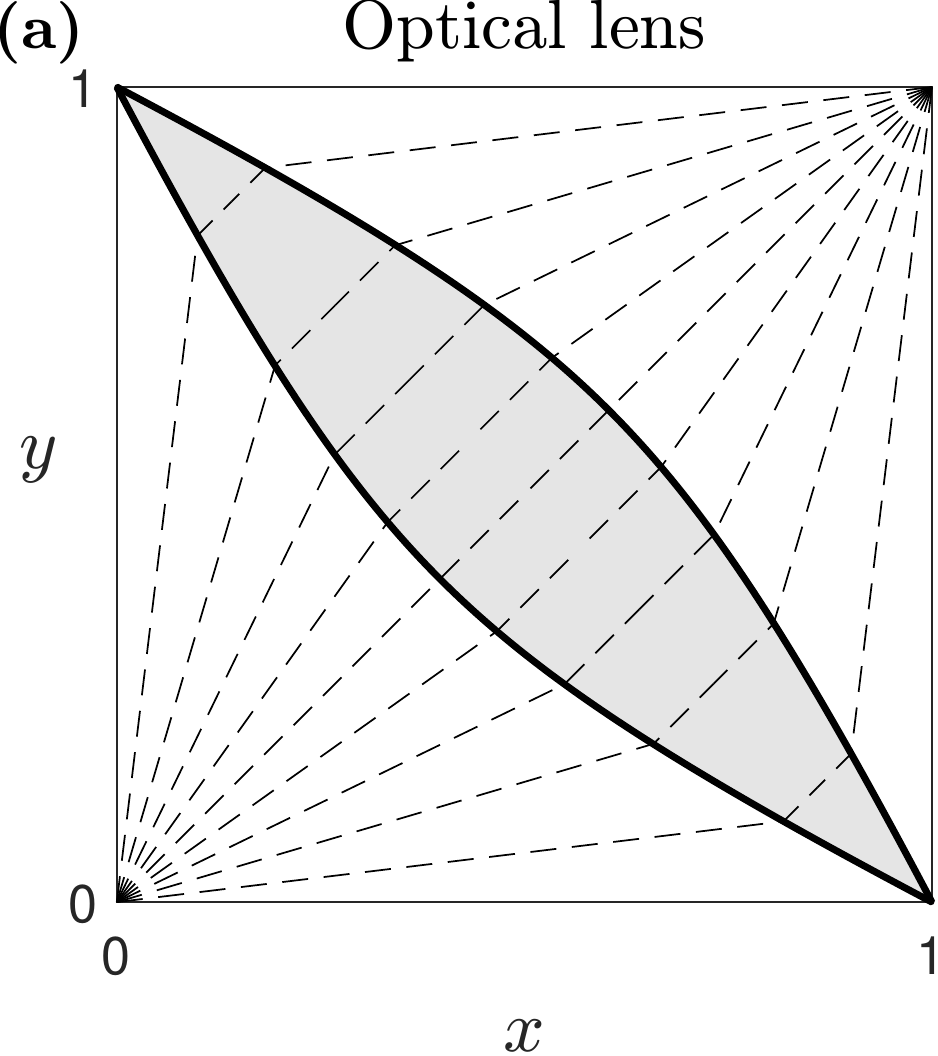}
		\hspace{0.03\linewidth}
		\includegraphics[height=0.5\linewidth]{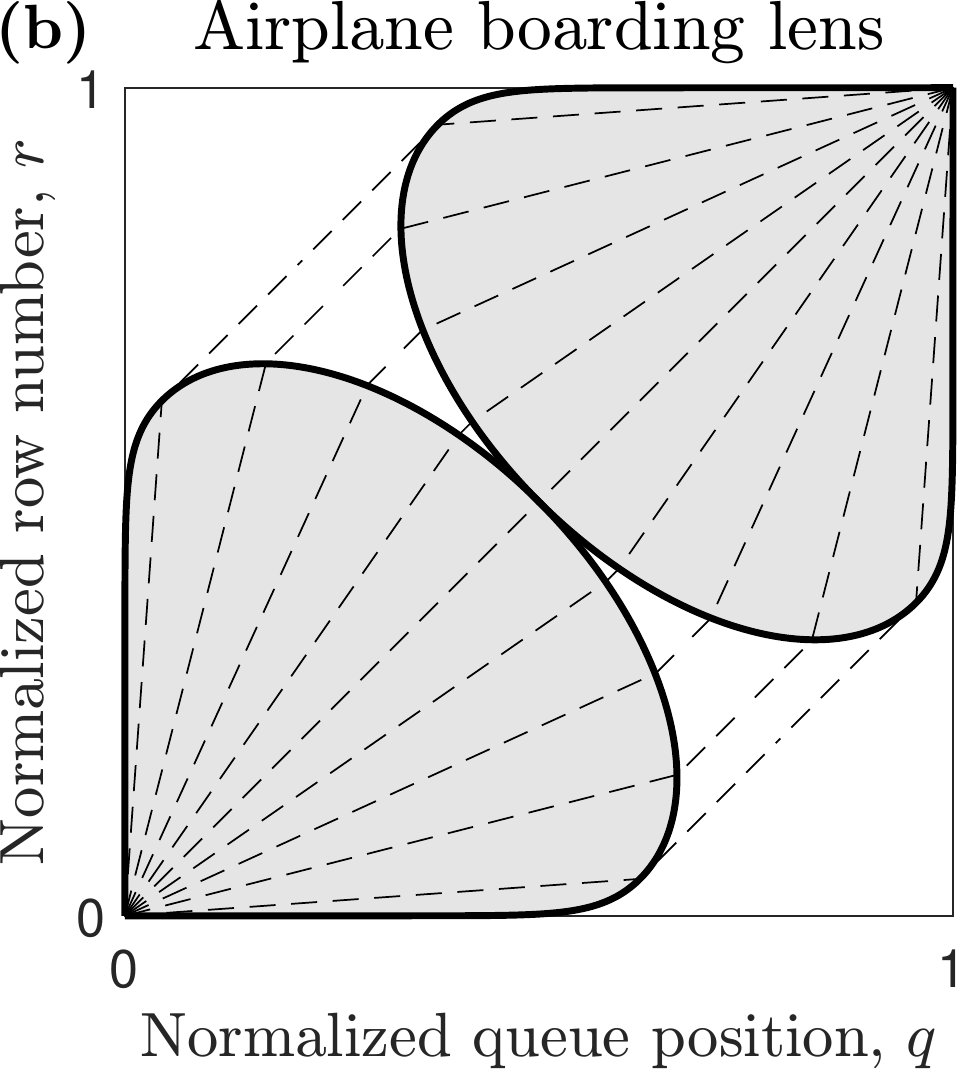}
		\caption{\label{fig:lenses} Lenses under the (a) Eucledian and (b) Minkowski metric, respectively. 
		For each point in the lens (shaded region), there is a a geodesic from $(0,0)$ to $(1,1)$ (dashed curves) passing through that point.
}
\end{figure}}

In a global economy, people, raw materials, products, power, and information are exchanged in large volumes, 
at high speeds, and over a range of scales in terms of distance.  At freight depots, arriving freight trains are
split and cars recombined into new trains according to destinations, using the available tracks \cite{Jaehn/Michaelis:2016}.  
Further examples range from  disk scheduling \cite{Bachmat:2014_ch3}, 
to internet traffic \cite{Larsson/Nilsson:2000}, 
to matching electricity production and consumption \cite{Han/Hilger/Mix/etal:2022}, 
to the travel pattern of cargo ships \cite{Kaluza/Kolzsch/Gastner/Blasius:2010}.

Bottlenecks arise dynamically during such logistics operations, and load and resources must be distributed optimally for a given infrastructure. 
While mitigation through re-routing is often possible in a network, this is not so in a queue. 

The case we will consider is the queue of passengers formed during airplane boarding.
We will explain the queue dynamics and its representation in simple diagrams. 
Optimal propagation is obtained through a lens-like construction in these diagrams. 
The lens structure is given by a closed-form expression 
and is a recipe for how to organize the queue at the gate so that slow passengers will not influence the boarding time.   
Note the similarity between the lens in \cref{fig:lenses}(b) and the standard optical lens in \cref{fig:lenses}(a): there are two media with different properties, and ``rays" follow straight paths inside each medium, but are ``refracted" at the interfaces. 
\vspace*{0.5ex}

\begin{figure*}[!t]
		\includegraphics[height=4.9cm]{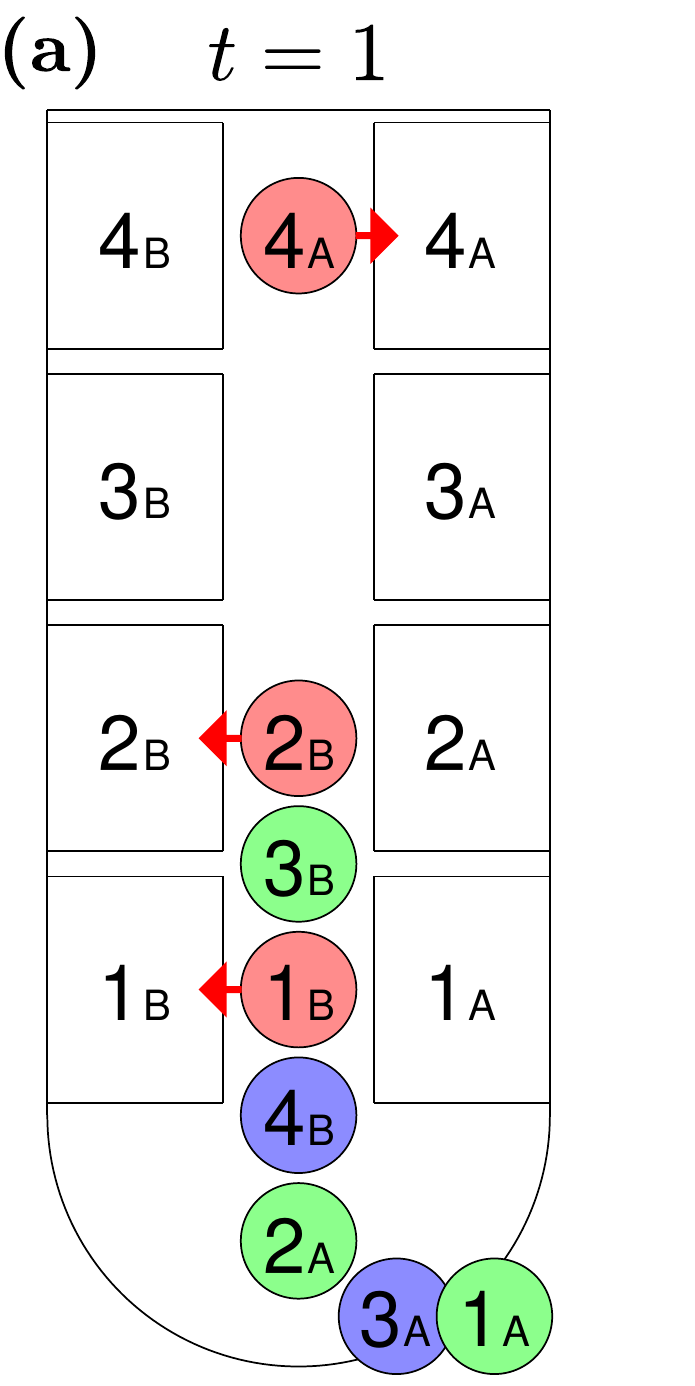}
		\includegraphics[height=4.9cm]{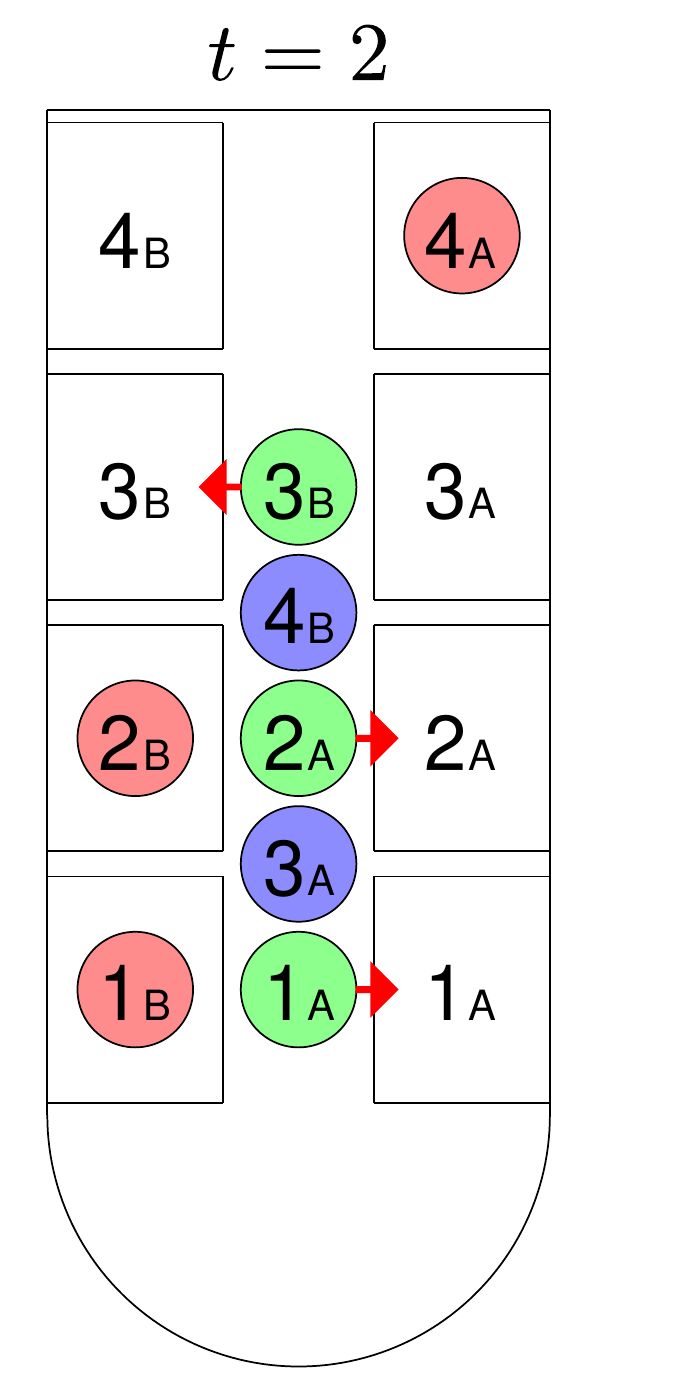}
		\includegraphics[height=4.9cm]{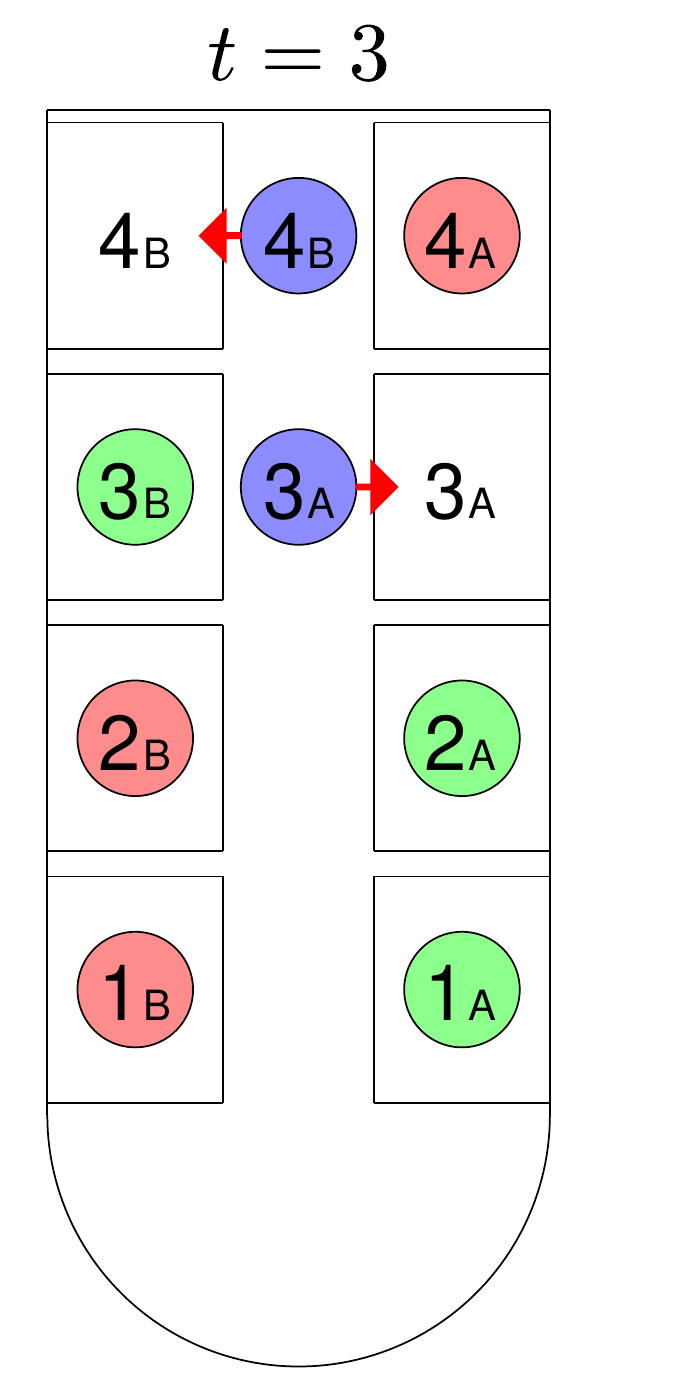}
		\includegraphics[height=4.9cm]{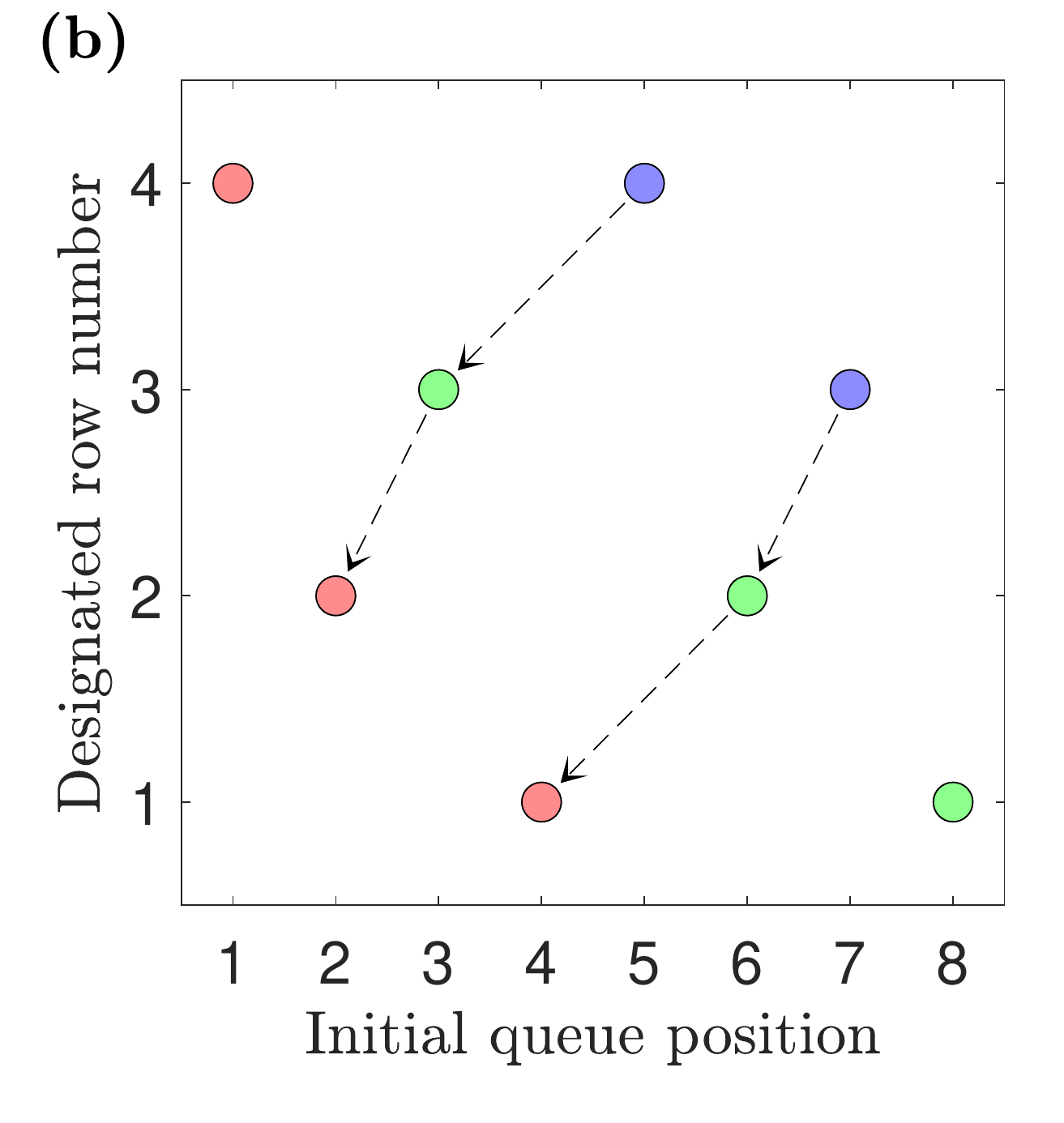}
		\includegraphics[height=4.9cm]{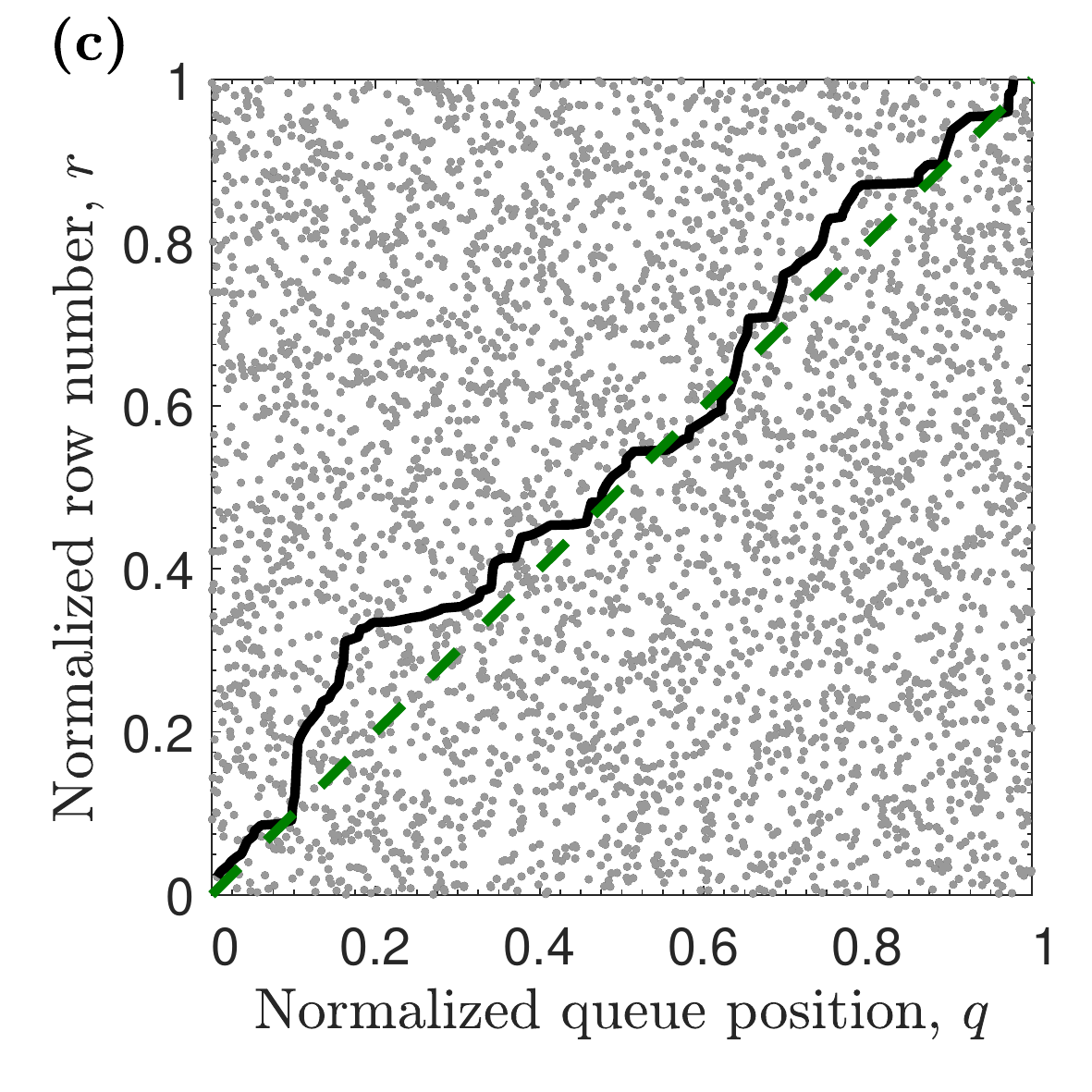}
	\caption{\label{fig:model} Boarding of an airplane.  Parts a and b show a small-scale example, with only $N=8$ passengers in 4 rows.  Passengers are indicated by circles, color coded by the time step they sit down.
	(a) The necessary 3 time steps during the boarding of one specific passenger order (queue structure).  Red arrows indicate passengers taking their seats.  The time for advancing along the aisle is assumed to be negligible compared with the aisle-clearing time. 	 
	(b) The initial queue in part a is represented in a diagram.  
	The abscissa represents initial queue positions, the ordinate designated row numbers.
	Arrows indicate blocking relations, as explained in the main text.
	(c) Diagram for a case with $N = 4000$ passengers.  Both axes have been normalized. 
	The black curve is the longest blocking chain, as	explained in the main text.  The green dashed line shows the limiting shape of the blocking chain as $ N \rightarrow \infty$.}
\end{figure*} 

{\em Boarding process.---}
Figure \ref{fig:model}(a) illustrates stages during boarding of an airplane.  Each passenger has a reserved seat, but cannot pass other passengers in the queue \nbb{who are still standing in the aisle.}
As the queue advances, passengers who have reached their row will be able to sit down. 
As a result of seat-taking in parallel, with $N$ passengers the total boarding time 
turns out to be $\sim N^{1/2}$ \cite{Bachmat/Berend/Sapir/Skiena/Stolyarov:2006}.

In \cref{fig:model}(b), the initial passenger queue is represented as points in a diagram. The abscissa is the location of the passenger in the queue, while the ordinate represents the designated row number. 
Passengers are often blocked from proceeding to their designated rows by those who have stopped to clear the aisle at a lower row number. Examples of such {\em blocking relations} are indicated by arrows in \cref{fig:model}(b). An example is passenger (3,3), who is blocked by (2,2).

Some passengers, like (7,3), are blocked by several other passengers. In general, a blocking passenger is often blocked by some other passenger.  This leads to a hierarchy of blockings, which can be represented as a chain connecting all passenger points involved.  For any passenger, {\em blocking chains} can be constructed,  usually ending very near the origin. The length of the longest chain determines the total boarding time  \cite{Bachmat:2014_ch3}.

In \cref{fig:model}(c), all $N=4000$ passengers are represented in the unit square in terms of their normalized queue $q$ and row location $r$. For simplicity the passengers are assumed to be paper thin, and a passenger $(q_1,r_1)$ blocks another passenger at $(q_2,r_2)$ if being in front in the queue $q_1 < q_2$ and having a lower row number $r_1 < r_2$. 
When $N\rightarrow\infty$, the longest blocking chain (black curve) is approaching a straight line from the first passenger taking seat near the front row to the last passenger sitting down in the back.

The blocking condition above can be shown to coincide with the past-future relation $ds^2>0$ of a positive-mass particle, under the Minkowski metric $ds^2=dt^2-dx^2$. 
The path between two events is the one that {\em maximizes} {(locally)} the proper time $\int ds$ between those events. Such maximized paths (geodesics) are straight lines.
	
In airplane boarding, the space-time coordinates $x,t$ are converted to light-like coordinates $q,r$ \nbb{(by a $45^{\circ}$ rotation)} with metric $ds^2=dqdr$. Hence, the asymptotic straight line in airplane boarding corresponds to the geodesic of a positive-mass particle in space-time geometry. It has been shown that the total boarding time $X_N$ is the proper time between $(0,0)$ and $(1,1)$, multiplied by a factor $2\tau\sqrt{N}$ when the passengers have equal aisle-clearing time $\tau$, and $N$ is large \cite{Myrheim:1978,Vershik/Kerov:1977, Logan/Shepp:1977, Bachmat/Berend/Sapir/Skiena/Stolyarov:2006}. \nbb{The same type of objects are also studied in the causal set program \cite{Bombelli/Lee/Meyer/Sorkin:1987, Surya:2012}.}

Different groups of passengers have different aisle-clearing time distributions, 
and can be classified as ``fast" and ``slow" depending on, e.g., the passengers' amount of carry-on luggage. The walking speed is assumed to be negligible compared to the aisle-clearing time.
Our policy problem is to find where we should place --- both in the queue and in the airplane ---
the group of slow passengers, i.e., which coordinates should slow passengers have in order to minimize their effect on boarding time. 

\begin{figure}[b]
	\includegraphics[width=0.49\linewidth]{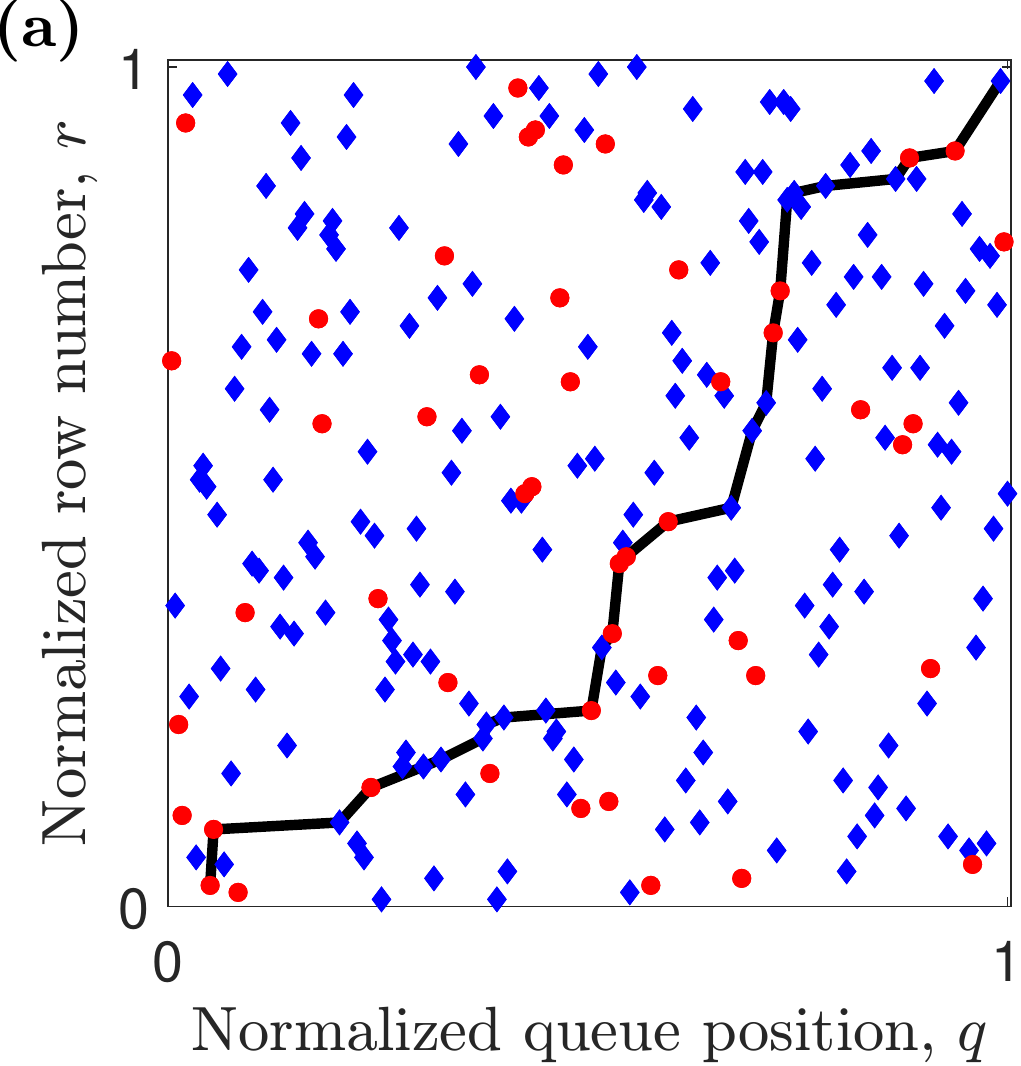}
	\includegraphics[width=0.49\linewidth]{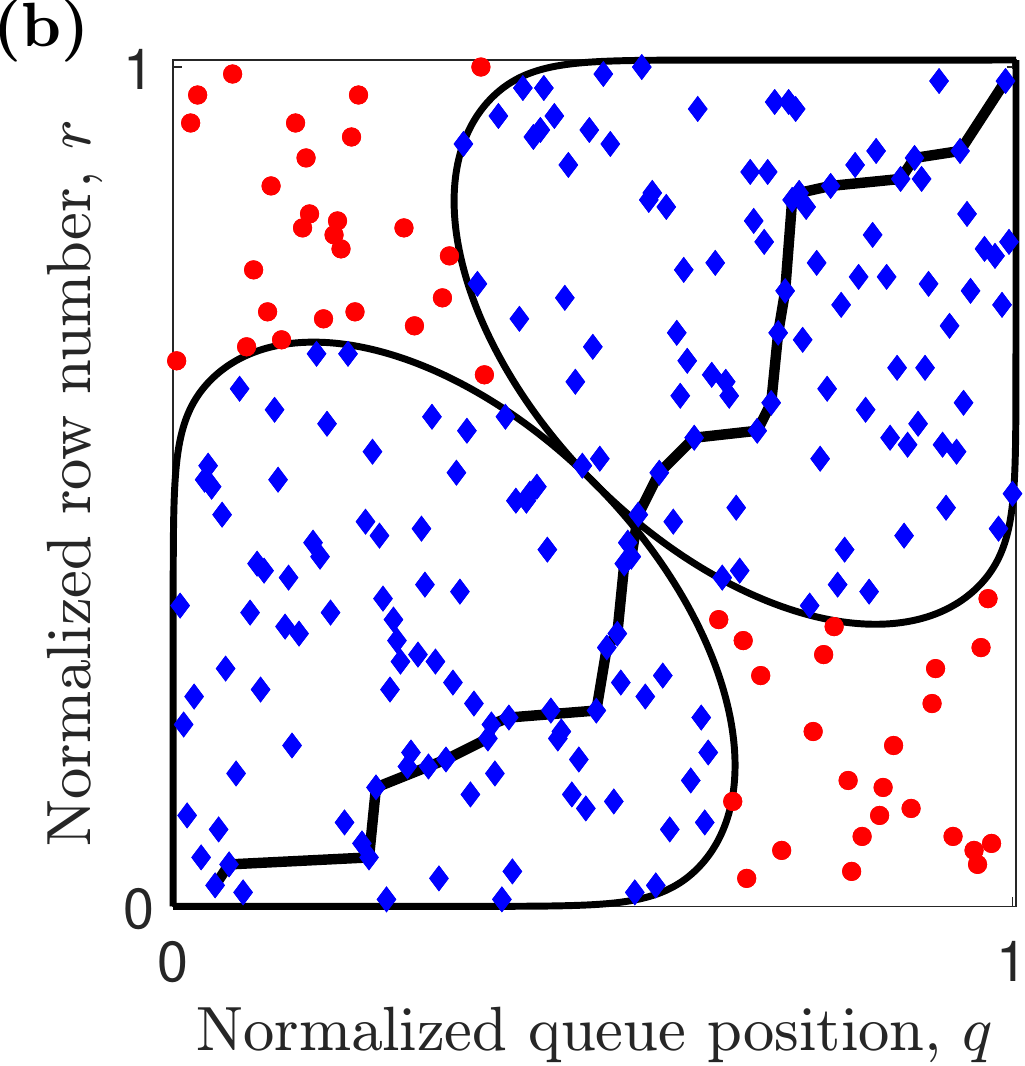}
	\caption{\label{fig:RAtoLENS} A fraction $p=20.8\%$ of the $N=240$ passengers are slow ({\color{red}$\bullet$}), with aisle-clearing time $T=2$ times larger than for the remaining fast passengers ({\color{blue}$\blacklozenge$}). 
		(a) Random boarding: passengers are randomly distributed in the $qr$-diagram, boarding takes $39$ time steps. 
		(b) Lens construction: the $qr$-location of the slow passengers are such that the boarding time remains the same as with fast passengers only ($28$ time steps).
		Note that the two blocking chains are slightly different.}
\end{figure} 

In \cref{fig:RAtoLENS}(a) the queue and row locations of both passenger groups are randomly distributed. 
Separation methods that reduce the boarding time already exist
\cite{Erland/Kaupuzs/Frette/Pugatch/Bachmat:2019,Erland/Kaupuzs/Steiner/Bachmat:2021,Bachmat/Erland/Jaehn/Neumann:2021}, but in \cref{fig:RAtoLENS}(b) the two groups of passengers are separated according to a lens construction that we will show completely eliminates the detrimental effect of the slow passengers.

According to the lens geometry in \cref{fig:RAtoLENS}(b), slow passengers who aim for rows in the back of the airplane \nbb{are {\em hidden} by placing them} in the first part of the queue ({\color{red}$\bullet$}~in upper left corner), while slow passengers who aim for the front rows should be placed in the back of the queue ({\color{red}$\bullet$}~in lower right corner). No slow passengers should have designated seats in the middle rows ($r\simeq 0.5$).\vspace*{0.5ex}

{\em Lens construction.---}
Here, we construct the lens shown in \cref{fig:lenses}(b) and \cref{fig:RAtoLENS}(b) for the asymptotic case when $N\rightarrow\infty$.  
The defining property of an optical converging lens is that there are (infinitely) many paths that solve Fermat's principle from a point $A$ to a focal point $B$. In Minkowski space there are many possible shapes for such lenses, however, we are not aware of any explicit constructions \footnote{\nbb{A preliminary form} of the lens shape has previously been documented as early work in a newsletter \cite{Bachmat/Erland:2020}, for which two of the present authors hold the copyright.}.

\begin{figure}[thb]
	\includegraphics[width=0.7\linewidth]{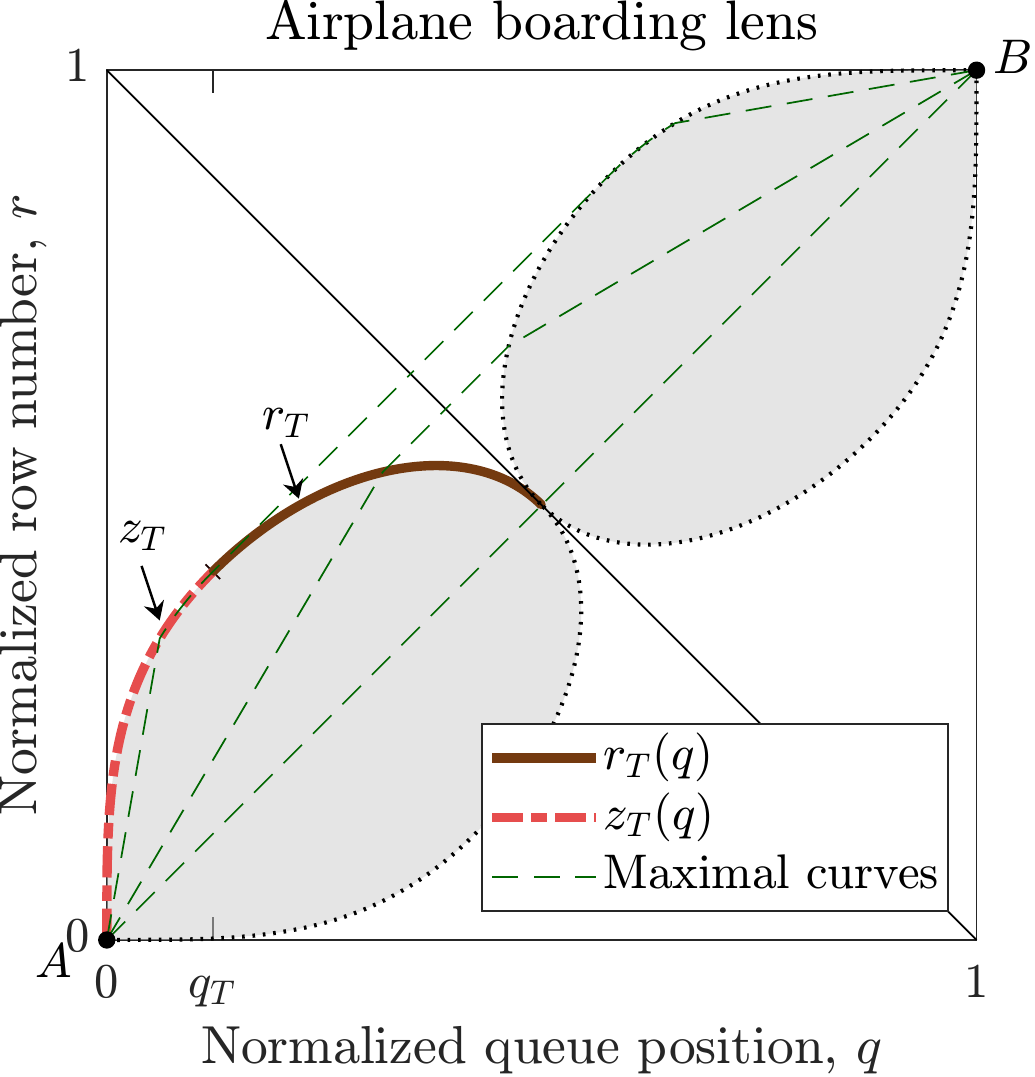}
	\caption{\label{fig:AB_Lensconstruction} 
		Lens construction: The lens is defined by the curves $r_T$ (solid line) and $z_T$ (dash-dotted line). Both curves are reflected through the diagonal and then through the antidiagonal (dotted lines). All geodesics emanating from $A$ can be extended to equal-length maximal curves (green dashed lines) under the Minkowski metric. Here $T=1.2$.}
\end{figure}

Consider the interval consisting of all points in light-like $q,r$ coordinates in the unit square, which are in the future of event $A(0,0)$ and in the past of event $B(1,1)$ in \cref{fig:AB_Lensconstruction}. We will assume that at each point in the interval, the metric can take either a standard or a scaled form. 
Let the lens $L$ be the set of points where the metric has the standard form $ds^2=dqdr$. Outside the lens, the scaled form $ds^2=T^2dqdr$ is taken, where $T>1$ can be considered as an index of refraction. In airplane boarding, the aisle-clearing time is 1 for the fast passengers inside $L$, and $T$ for the slow passengers outside $L$. 
 
The lens property we are aiming for is that any time-like geodesic leaving $A$ will be part of a maximal proper time curve from $A$ to $B$ with respect to $\int ds$. Thus, free falling positive-mass particles leaving $A$ will refocus at $B$ with synchronized proper time, and in airplane boarding there will be infinitely many blocking chains of the same length.

Moreover, we search for solutions where the total boarding time is the same as if all passengers were fast.  
Thus, the proper time of maximal curves passing through areas outside the lens must be equal to the proper time of the straight-line maximal curve between $A$ and $B$ with the standard form metric. 
We call a lens with this property a {\em critical lens}.

We first consider the triangle of the unit square in \cref{fig:AB_Lensconstruction}, consisting of the area between the line $q=0$, the diagonal, and the anti-diagonal. 
The lens is then extended to the entire unit square by reflecting through the diagonal and then through the anti-diagonal. 

In optics, spherical lenses induce (spherical) aberrations. For lenses shaped according to Cartesian ovals (as in \cref{fig:lenses}(a)), there are no such aberrations \cite{Descartes:1637_App3,Descartes:1925,Maesumi:1992}.
Hence, we start with Descartes' construction of Cartesian oval lenses for the degenerate case where the source is $A(0,0)$ and the target the anti-diagonal $r=1-q$ (focus at infinity) adapted to Minkowski geometry. 
Referring to \cref{fig:AB_Lensconstruction}, we initially assume that the lens border is defined by a function $r_T(q)$ and that the geodesic (middle dashed curve) will be a line from the origin to a point $(q,r_T(q))$, of proper time $\sqrt{q r_T(q)}$ inside the lens. Outside the lens the maximal curve must, due to symmetry, continue towards the anti-diagonal as a slope-1 line, having proper time $T(1-q-r_T(q))/2$. 
Imposing that the total proper time must be the same as when the metric is unscaled (geodesic along the diagonal in \cref{fig:AB_Lensconstruction}), leads to a degenerate Descartes equation 
\begin{equation}\label{eq:Equallength}
	\sqrt{qr_T(q)} + \frac{T}{2}(1-q-r_T(q)) = \frac{1}{2}, 
\end{equation}
with the explicit solution when $q\in [0,1/2]$
\begin{align}
		r_T(q)=&\frac{1}{T^2}\left[\sqrt{q} + \sqrt{T^2 -T -q(T^2-1)}\right]^2.
\end{align}

However, the assumption that the slope-1 segment from $(q,r_T(q))$ to the anti-diagonal will be outside the lens only holds if $r_T'(q)\leq 1$, which is true only in the range $q\in [q_T, 1/2]$ with 
\begin{equation}
	\label{x-def}
	q_T=\frac{T-\sqrt{T^2-1}}{2(T+1)}.
\end{equation}

For $q<q_T$, a geodesic curve from $A$ (like the leftmost green dashed line in \cref{fig:AB_Lensconstruction}) cannot be refracted in such a way that it will be perpendicular to the anti-diagonal. Thus, in this interval 
we ``weld" to the Descartes curve $r_T(q)$ in \cref{eq:Equallength}, a curve of the form
\begin{equation}
	\label{z-def}
	z_T(q)=c_Tq^{\alpha_T}, \qquad \alpha_T =(T-\sqrt{T^2-1})^2.
\end{equation}
Such curves have the key property that the length of a curve segment between the origin and a point $(q,z_T(q))$ (using the $T$-scaled metric along the lens border) equals the length of a straight line segment between the same points inside the lens (standard metric). 

We choose $c_T$ so that the curve $z_T$ will meet the Descartes curve at $(q_T,r_T(q_T))$. This occurs when
\begin{equation}\label{c-def}
	c_T=q_T^{1-\alpha_T}/\alpha_T.
\end{equation}
Note that $z_T'(q_T)=r_T'(q_T)=1$, so the welding is smooth.

All causal curve geodesics emanating from $A$ with end point in $B$ are represented by the three cases in \cref{fig:AB_Lensconstruction}. By construction, they have equal proper time. 
Note that blocking chains in airplane boarding are \emph{global} optima, while positive-mass particles follow \emph{locally} optimal paths. Hence, geodesics cannot go along the $z_T$-part of the lens border in \cref{fig:AB_Lensconstruction}, \nbb{unless they are guided}, and the lens property is only maintained for particles meeting the lens border at the $r_T$-part, after emanating from $A$.

The constructed lens is {\em minimal} in the sense that if any of the fast passengers inside the lens is replaced by a slow passenger, the boarding time will increase. Thus, all points in the lens are part of a maximal curve. Without proof, we also conjecture that the constructed lens is minimal in the sense that the area of the lens (the fast passengers) is minimized. \vspace*{0.5ex}

{\em Maximal number of slow passengers.---}  
The area of the critical lens can be computed explicitly from the expressions for $r_T$ and $z_T$. For $T=2$, up to 26.5\% slow passengers will lead to the same asymptotic boarding time as for only fast passengers when $N\rightarrow\infty$.
The corresponding asymptotic boarding time with random boarding (as in \cref{fig:RAtoLENS}(a)) is 39\% higher.

\begin{figure}[tb]
	\includegraphics[width=0.5\linewidth]{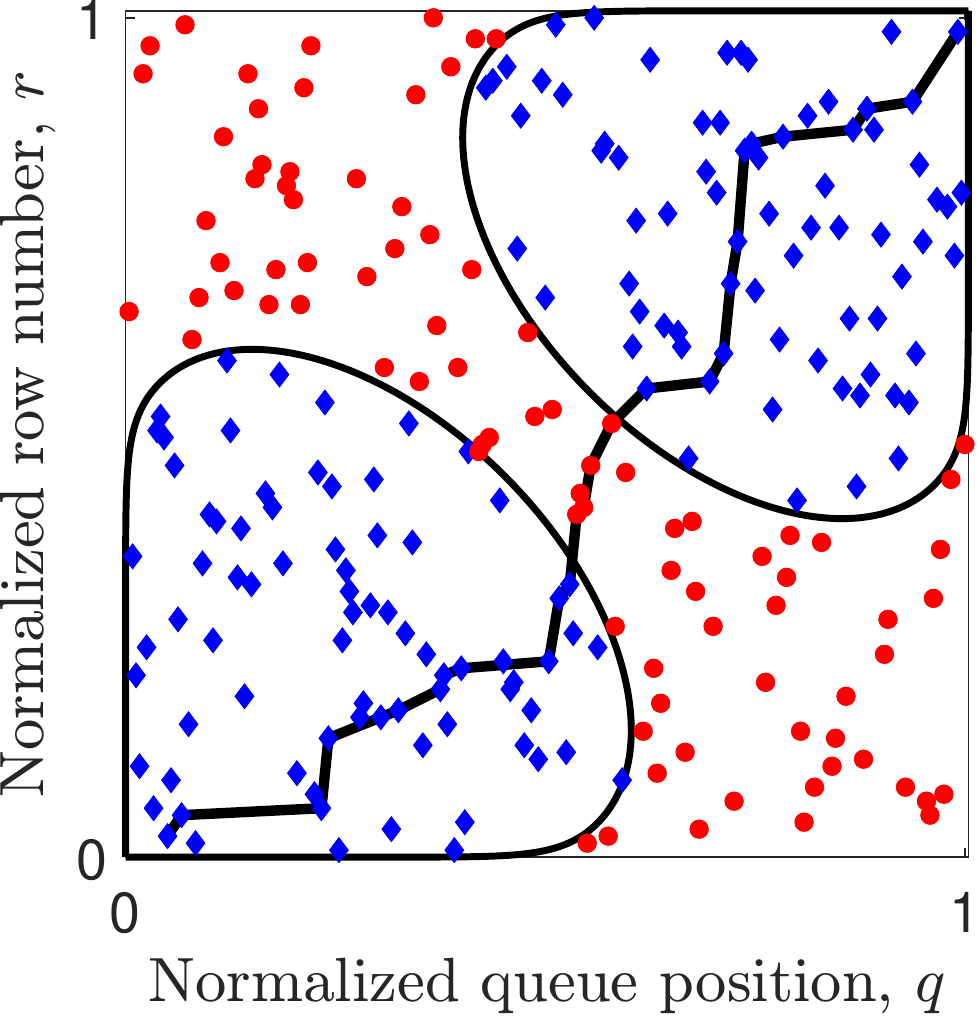}
	\caption{\label{fig:Lens_subsuper} A subcritical lens is obtained by scaling, $c=0.9$, $T=2$ and $N=240$. The lens property is maintained, and the asymptotic fraction of slow passengers is increased from 26.5\% to 40.5\%. The asymptotic boarding time increase by 10\% (and here with $N=240$, from 28 to 32 time steps, see \cref{fig:RAtoLENS}(b)).
	}
\end{figure} 
We can also use the above to construct subcritical lenses as in \cref{fig:Lens_subsuper}. These are useful if the proportion of fast  passengers is smaller than that required by the critical lens.
Subcritical lenses can be constructed by scaling the lens 
by a constant $c<1$. Thus, $r_T(q)$ is substituted by $\tilde{r}_T(q)=cr_T(q/c)$ for $q \in (cq_T,c/2)$ and likewise with $z_T$. 
There will still be maximal curves passing through every point of the lens, but their length (i.e., the boarding time) increases with a factor $c+(1-c)T$. 
The proportion of fast passengers is reduced by a factor $c^2$. \vspace*{0.5ex}

\begin{figure}[tb]
	\includegraphics[width=0.49\linewidth]{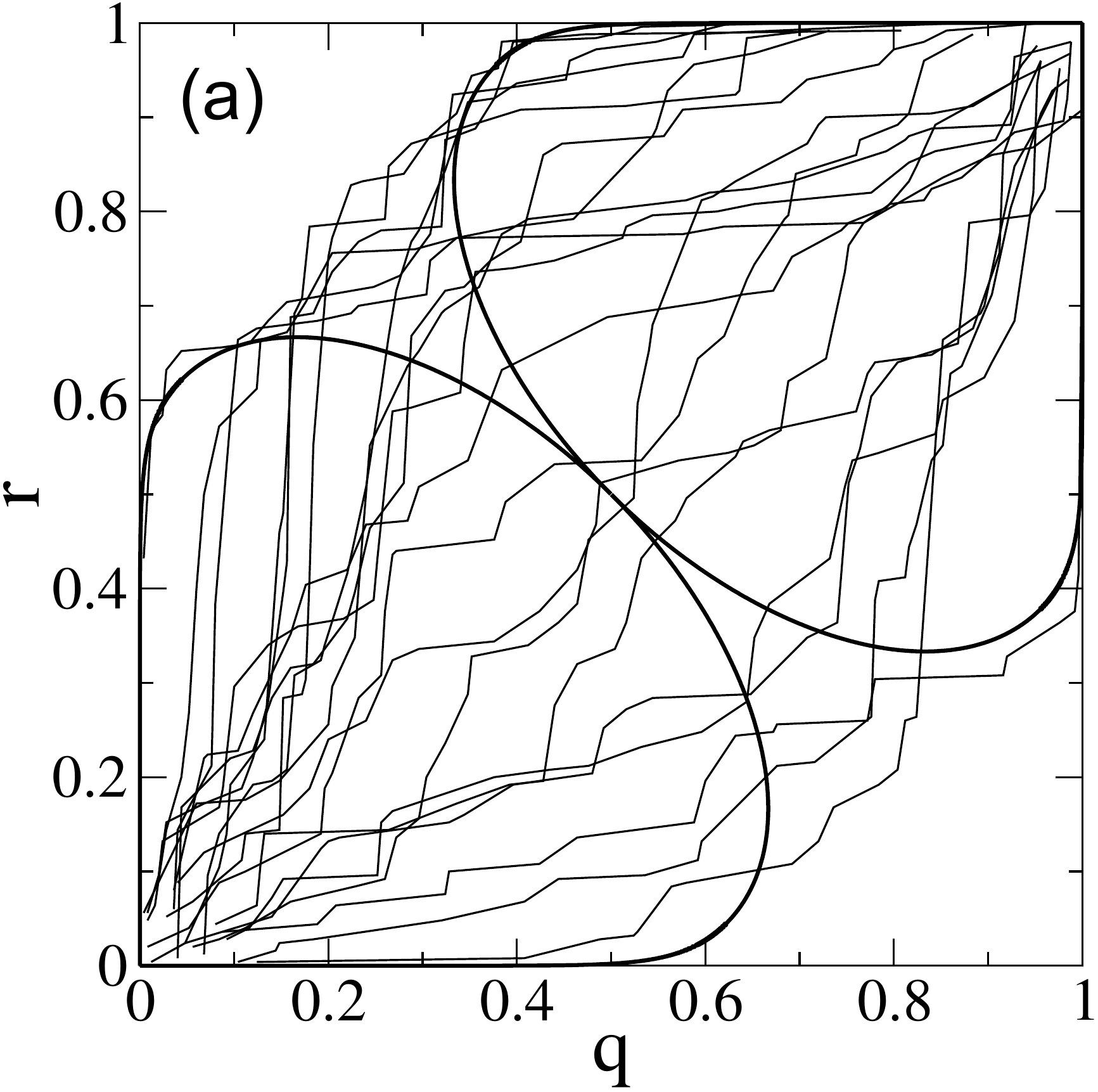}
	\includegraphics[width=0.49\linewidth]{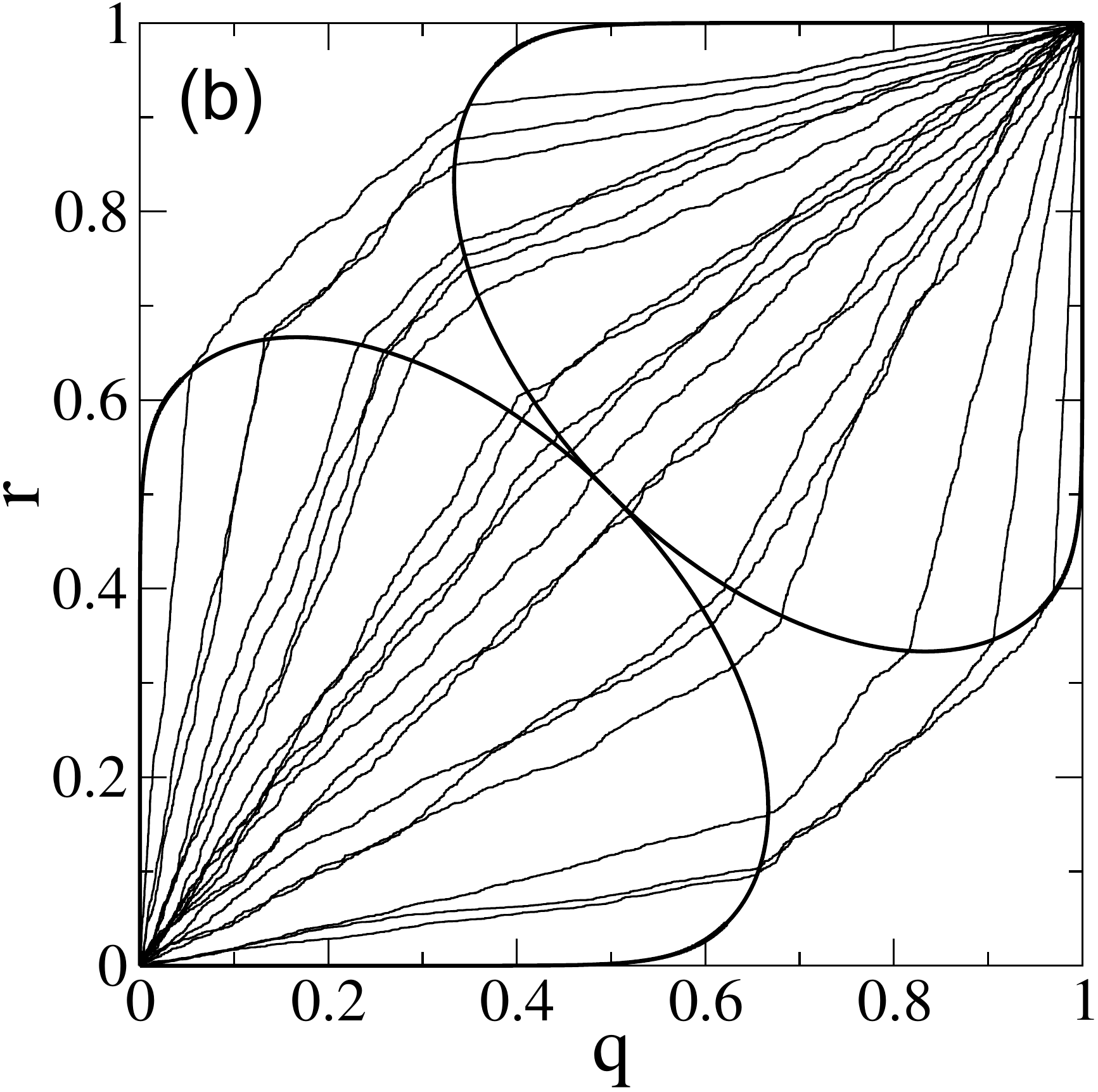}
	\caption{\label{fig:lens_Ndep} Blocking chains from several different queue realizations. Increasing number of passengers $N$ gives more regular blocking paths. $T=2$. (a) $N=250$. (b) $N=10^7$.}
\end{figure}

{\em Lens convergence rates.---} 
Figure \ref{fig:RAtoLENS}(b) shows one blocking chain --- among many possible of equal or near-equal length --- that pass through the boarding-queue lens. As $N$ increases, the blocking chains \nbb{straighten out}, as demonstrated in \cref{fig:lens_Ndep}, but the convergence is slow.

\begin{figure}[b]
	\includegraphics[width=1\linewidth]{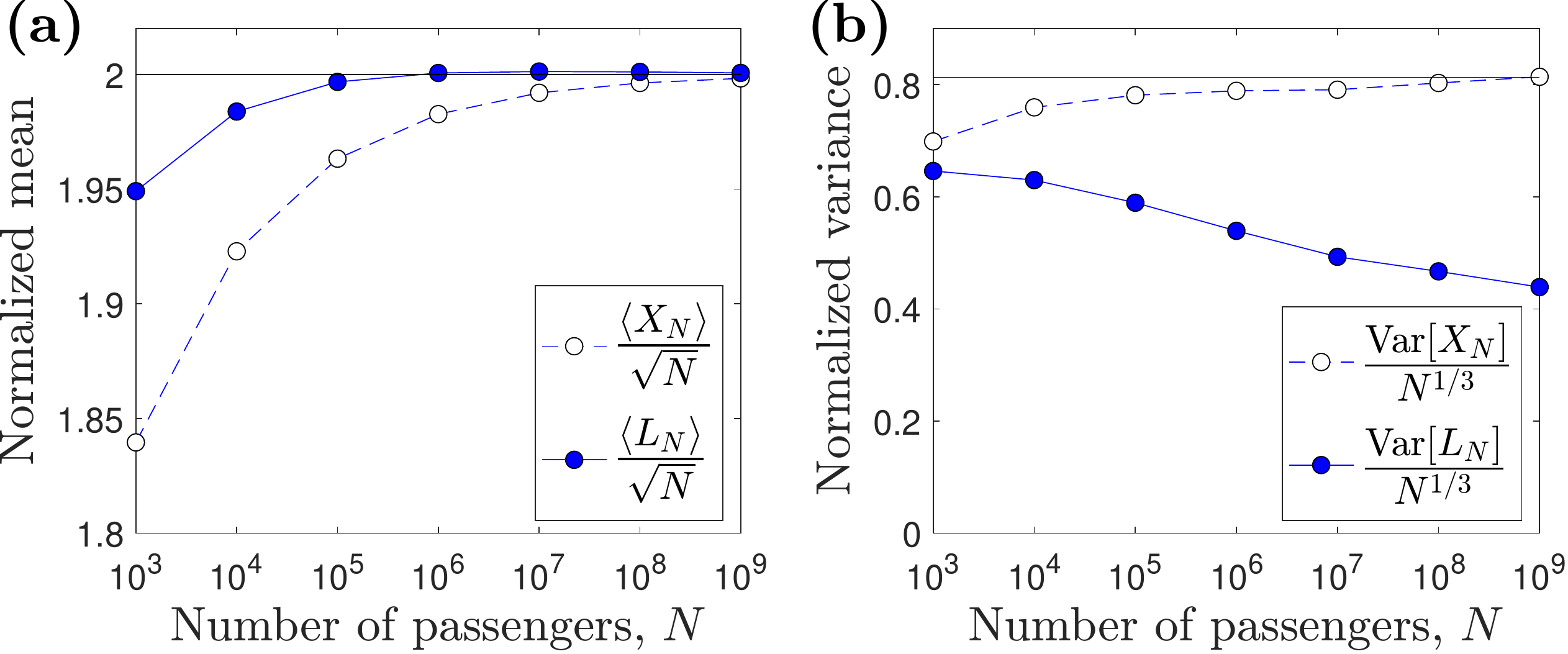}
	\caption{\label{fig:varreduction} (a) Mean and (b) variance of finite-$N$ boarding times. The lens construction ($L_N$, $T=2$) has faster convergence with less variance than with fast passengers only ($X_N$).}
\end{figure}

Nevertheless, the rate of convergence to the asymptotic total boarding time is faster with the lens construction ($L_N$) than with only fast passengers ($X_N$). With the lens there are more competing chains of near-equal length, and the length of the maximal chain $L_N$ tends to be longer and closer to the asymptotic estimate than $X_N$ (see \cref{fig:varreduction}). 
\nbb{While the fluctuations of the normalized boarding time $X_N/\sqrt{N}$ will have a GUE Tracy-Widom distribution, with a \emph{negative} leading order correction $N^{-1/6}$ \cite{Baik/Deift/Johansson:1999, Bachmat/Berend/Sapir/Skiena/Stolyarov:2006}, the extensive simulations in \cref{fig:varreduction}(a) confirm that the lens has a \emph{positive} leading order correction for large $N$ \cite{Bachmat:2014_ch3}. The results in \cref{fig:varreduction}(b) indicate that the variance of $L_N$ is of a smaller order than $N^{1/3}$ of $X_N$.}

{\em Extensions and further applications.---}
Note that lenses with respect to proper time also exist in other models, e.g., 
\nbb{any point in Anti-deSitter space has a conjugate point, and the points are connected by infinitely many geodesics}
\cite{Penrose:1972}.
\nbb{Further results and applications will be reported elsewhere},
including lenses with precise convergence rate estimates; 
lenses in higher dimensions (with increased fraction of hidden slow passengers); 
lenses for practical implementation in airplane boarding, \nbb{accounting for the queue displacement when passengers} are no longer infinitely thin (as shown in \cref{fig:model}(a)); 
polynuclear growth models (designing flat surfaces) \cite{Prahofer/Spohn:2000, Ferrari:2004png};
\nbb{lens constructions for both acoustic and electromagnetic waves in metamaterials}; and more.


%

\end{document}